\documentclass[usenatbib]{mnras}
\usepackage{graphicx, amssymb, aas_macros, natbib, bm}
\usepackage{ marvosym }
\usepackage{mathtools}
\usepackage{enumitem}
\usepackage{fancybox}
\usepackage{lipsum}  
\usepackage{gensymb}
\usepackage{calc}
\usepackage{csquotes}
\usepackage{xcolor}
\usepackage{amsmath}
\usepackage{multicol}
\usepackage{multirow}
\usepackage{orcidlink}

\graphicspath{{plots/}}

\providecommand{\pgfsyspdfmark}[3]{}
\usepackage[english]{babel}
\defcitealias{Baxter22}{B22}

\newcommand{\mtwo}{M_{\rm{200}}}
\newcommand{\rtwo}{R_{\rm{200}}}

\newcommand{\mhalo}{{M}_{\rm{halo}}}

\newcommand{\mstar}{{M}_{\star}}

\newcommand{\msun}{{\rm M}_{\odot}}

\newcommand{\lt}{<}
\newcommand{\gt}{>}

\newcommand{\oii}{\hbox{\sc [O\,ii]}}

\DeclareRobustCommand{\ion}[2]{%
\relax\ifmmode
\ifx\testbx\f@series
{\mathbf{#1\,\mathsc{#2}}}\else
{\mathrm{#1\,\mathsc{#2}}}\fi
\else\textup{#1\,{\mdseries\textsc{#2}}}%
\fi}

\title[Environmental Quenching in Clusters at $z \gtrsim 1$]
{When the Well Runs Dry: Modeling Environmental Quenching of High-mass Satellites in Massive Clusters at \boldmath$z \gtrsim 1$}

\author[Baxter et al.]
{Devontae C. Baxter${\orcidlink{0000-0002-8209-2783}}$,$^1$\thanks{$\!\!$e-mail: dcbaxter@ucsd.edu}\thanks{$\!\!$LSSTC DSFP Fellow}
M. C. Cooper${\orcidlink{0000-0003-1371-6019}}$,$^1$
Michael L. Balogh${\orcidlink{0000-0003-4849-9536}}$,$^{2,3}$
\newauthor
Gregory H. Rudnick${\orcidlink{0000-0001-5851-1856}}$,$^{4}$
Gabriella De Lucia${\orcidlink{0000-0002-6220-9104}}$,$^{5}$
Ricardo Demarco${\orcidlink{0000-0003-3921-2177}}$,$^{6}$
\newauthor
Alexis Finoguenov${\orcidlink{0000-0002-4606-5403}}$,$^{7}$
Ben Forrest${\orcidlink{0000-0001-6003-0541}}$,$^{8}$
Adam Muzzin${\orcidlink{0000-0002-9330-9108}}$,$^{9}$
\newauthor
Andrew M. M. Reeves${\orcidlink{0000-0003-2618-6408}}$,$^{2, 3}$
Florian Sarron${\orcidlink{0000-0001-8376-0360}}$,$^{10}$
Benedetta Vulcani${\orcidlink{0000-0003-0980-1499}}$,$^{11}$
\newauthor
Gillian Wilson$\orcidlink{0000-0002-6572-7089}$,$^{12}$
Dennis Zaritsky${\orcidlink{0000-0002-5177-727x}}$$^{13}$  \\ \\
\newline \noindent {\normalsize \it Affiliations are listed at the end of the paper}
}

\begin{document}

\pagerange{\pageref{firstpage}--\pageref{lastpage}} 
\pubyear{2022}

\maketitle

\label{firstpage}
\begin{abstract}

We explore models of massive ($\gt 10^{10}~\msun$) satellite quenching in massive clusters at $z\gtrsim1$ using an MCMC framework, focusing on two primary parameters: $R_{\rm quench}$ (the host-centric radius at which quenching begins) and $\tau_{\rm quench}$ (the timescale upon which a satellite quenches after crossing $R_{\rm quench}$). Our MCMC analysis shows two local maxima in the 1D posterior probability distribution of $R_{\rm quench}$ at approximately $0.25$ and $1.0~\rtwo$. 
Analyzing four distinct solutions in the $\tau_{\rm quench}$-$R_{\rm quench}$ parameter space, nearly all of which yield quiescent fractions consistent with observational data from the GOGREEN survey, we investigate whether these solutions represent distinct quenching pathways and find that they can be separated between \textquote{starvation} and \textquote{core quenching} scenarios. 
The starvation pathway is characterized by quenching timescales that are roughly consistent with the total cold gas (H$_{2}$+H{\scriptsize I}) depletion timescale at intermediate $z$, while core quenching is characterized by satellites with relatively high line-of-sight velocities that quench on short timescales ($\sim 0.25$ Gyr) after reaching the inner region of the cluster ($\lt 0.30~\rtwo$).  
Lastly, we break the degeneracy between these solutions by comparing the observed properties of transition galaxies from the GOGREEN survey.
We conclude that only the \textquote{starvation} pathway is consistent with the projected phase-space distribution and relative abundance of transition galaxies at $z \sim 1$.  
However, we acknowledge that ram pressure might contribute as a secondary quenching mechanism.

\end{abstract}


\section{Introduction}
\label{sec:intro} 

Environmental studies in the local Universe and extending out to $z\sim2$ have found that galaxies that are members of massive galaxy groups and clusters – i.e. satellites – are more likely to be passive (or quenched) relative to their counterparts of similar mass in the low-density field \citep{Oemler74, Dressler80, Balogh97, Gomez03, Baldry06, Cooper06, Cooper07, Cooper10a, Guo17, LeeBrown17, Ji18, Lemaux19, Pintos19, Shi21, McConachie21}. It has long been understood that satellite galaxies – by virtue of their environment – are uniquely subject to a variety of environmental quenching mechanisms \citep[e.g.][]{Baldry06, Peng10, Peng12} that suppress star formation by way of (\emph{i}) gas depletion without replenishment or (\emph{ii}) stripping and removal of cold gas (i.e. the fuel for star formation). Two of the leading environmental quenching candidates that satisfy these conditions include \textquote{starvation} \citep{Larson80, Bekki02, Kawata08} -- the \emph{slow} depletion of cold gas in the absence of cosmological accretion after a galaxy becomes a satellite of a massive host -- and \textquote{ram-pressure stripping} \citep[RPS;][]{GG72, Abadi99, Poggianti17} -- the \emph{rapid} removal of cold gas from the interstellar medium of a satellite as it moves through the dense intra-group or intracluster medium permeating the host halo. Other potential environmental quenching mechanisms include gravitationally-driven processes such as tidal stripping \citep{Merritt83, Moore99, Gnedin03}, mergers \citep{L&H88, Makino97, Gottlobrer01}, and galaxy harassment via high-speed impulsive encounters \citep{Farouki81, Moore96, Moore98} -- as well as \textquote{outflow-based} processes such as overconsumption \citep{McGee14, Balogh16}. 

Although these mechanisms primarily impact fully accreted satellites, several studies have shown that galaxies can undergo \textquote{group pre-processing} \citep{Fujita04, DeLucia12, Wetzel15, Bianconi18, Sarron19, Werner22}, wherein they quench within a more massive halo prior to becoming a satellite of the final group or cluster. 
Thus, understanding the dominant driver of environmental quenching is challenging, as it entails making assumptions regarding the relative contributions of pre-processed satellite galaxies to the observed quiescent fraction. 
Moreover, an additional challenge is that environment-independent quenching processes \citep[often referred to as \textquote{mass quenching} or \textquote{self-quenching},][]{Peng10} may be dominant, particularly for massive galaxies \citep[e.g.][]{Tacchella15, Reeves21, Werner22, Syeda23}.
Such mechanisms, including feedback from star formation \citep{Oppenheimer06, Ceverino09}, supernovae \citep{Springel05, Lagos13}, and active galactic nuclei \citep{DiMatteo05, Croton06, hopkins06}, are capable of quenching galaxies prior to them becoming fully incorporated into a galaxy group or cluster.
%

At present, our current understanding of the dominant quenching mechanism driving environmental quenching in galaxy groups and clusters is largely limited to the very local ($z \lt 0.1$) Universe \citep[e.g.][]{DeLucia12, Wetzel13, Hirschmann14, Wheeler14, Fham15, Fham16, Fham18, Davies16, Pallero19, RodriguezWimberly19, Baxter21}. 
In fact, our best cosmological models routinely fail to reproduce the observed fraction of quenched satellites as a function of stellar mass beyond the local Universe, signaling that our current prescriptions for environmental quenching are incomplete 
at intermediate and high redshift \citep[e.g.][]{Guo2010, Hirschmann14,DeLucia19, Xie20, Donnari21, Kukstas23}.

In our recent work \citep[][hereafter \citetalias{Baxter22}]{Baxter22}, we built upon previous efforts to constrain the dominant quenching mechanism in massive clusters at $z \sim 1$ \citep[e.g.][]{Muzzin14, Balogh16, Foltz18} by constraining the timescale ($\tau_{\rm quench}$) upon which satellite quenching proceeds following infall.
Given that different mechanisms operate on distinct timescales, knowledge of $\tau_{\rm quench}$ at a given epoch can aid in distinguishing the underlying quenching mechanism at play \citep[e.g.][]{Wetzel14, Fham15, Wright19, Park22}.
In \citetalias{Baxter22}, we accomplish this by developing an infall-based environmental quenching model -- with prescriptions for \textquote{field quenching} (i.e.~self-quenching in the field) and \textquote{pre-processing} -- that infers the quenching timescale consistent with the observed satellite quiescent fraction as a function of stellar mass as measured in $14$ massive clusters ($\mhalo = 10^{14-15}~\msun$) from the GOGREEN survey \citep{Balogh21} -- the hitherto largest and most comprehensive spectroscopic and multi-passband photometric cluster and group survey at $z \gtrsim 1$. 
Many of the conclusions drawn in \citetalias{Baxter22} are consistent with results from previous GOGREEN studies \citep{Webb20, Reeves21, McNab21}, including that (\emph{i}) the majority of massive galaxies ($\mstar \gtrsim 10^{10.5}~\msun$) quench before they become cluster members and (\emph{ii}) low-mass galaxies ($\mstar \lesssim 10^{10.5}~\msun$) are preferentially quenched after infall.
In addition, the analysis presented in \citetalias{Baxter22} finds that the satellite quenching timescale at $z \sim 1$ is in good agreement with the estimated cold gas (H{\scriptsize I}+${\rm H}_2$) depletion timescale, suggesting that starvation may be the dominant quenching mechanism within GOGREEN clusters.

While the modeling from \citetalias{Baxter22} suggests that the inferred satellite quenching timescale in massive clusters is consistent with starvation being the dominant driver of environmental quenching at $z < 2$, there is a wealth of literature showing that RPS is an active process in cluster environments in the nearby Universe \citep[][]{Yagi07, Boselli16b, Gavazzi18, Moretti18, Vulcani18, Poggianti19, Gullieuszik20, Luber22}. 
Moreover, recent observational studies find direct evidence of satellites in clusters at $z \sim 0.7-1.6$ suffering from RPS \citep{Boselli19, Noble19, Matharu21, Cramer23}, while simulations find that RPS should be effective in cluster environments up to $z \sim 2$ \citep[see review from][]{Boselli22}. 
Given that the efficiency of RPS depends directly on the density of the intracluster medium (ICM) -- i.e. higher near the core of a cluster -- we generalize the environmental quenching model developed in \citetalias{Baxter22} to include the radius at which quenching begins ($R_{\rm quench}$) as a free parameter.
In addition, we explore model results regarding where within the cluster and with what velocity satellites quench. 
These modifications permit the exploration of potentially distinct quenching pathways, by no longer assuming that environmental quenching begins immediately after crossing $\rtwo$, allowing us to test whether or not the main conclusion drawn in \citetalias{Baxter22} -- i.e.~whether starvation is the dominant quenching pathway at $z < 2$ -- is robust to changes in our modeling regarding where in a cluster environmental quenching becomes effective.
%

In \S\ref{sec:GOGREEN} and \S\ref{sec:TNG} of this work, we describe our observed galaxy cluster sample and our simulated satellite population, respectively -- leaving details regarding cluster membership criteria to \citetalias{Baxter22}. In \S\ref{sec:Model}, we describe our updated environmental quenching model, with the results from our MCMC analysis and comparison of model predictions with observed properties of transition galaxies presented in \S\ref{sec:Results}. In \S\ref{sec:Discussion}, we discuss our procedure for isolating distinct quenching pathways and contextualize our results with respect to previous studies at $z \sim 1$. Finally, in \S\ref{sec:Conclusion} we summarize our investigation and present our conclusions.
When necessary, we adopt a flat $\Lambda$CDM cosmology with $H_{0} = 70~{\rm km}~{\rm s}^{-1}~{\rm Mpc}^{-1}$ and $\Omega_{m}$ = 0.3 as well as a \citet{Chabrier03} initial mass function. All magnitudes are on the AB system \citep{OkeGunn83}.


\section{Observed Cluster Sample}
\label{sec:GOGREEN}

\subsection{GOGREEN and GCLASS Cluster Sample}
\label{subsec:2.1}

We select our cluster sample from the Gemini CLuster Astrophysics Spectroscopic Survey (GCLASS) and the Gemini Observations of Galaxies in Rich Early ENvironments (GOGREEN) surveys \citep{Muzzin12, Balogh17, Balogh21}.\footnote{\href{http://gogreensurvey.ca/data-releases/data-packages/gogreen-and-gclass-first-data-release/}{http://gogreensurvey.ca/data-releases/data-packages/gogreen-and-gclass-first-data-release/}} The main focus of these surveys is to study galaxy evolution in high-density environments by combining deep, multi-wavelength photometry with extensive Gemini/GMOS \citep{Hook04} spectroscopy of galaxies in 26 overdense systems over a redshift range of $0.867 \lt z \lt 1.461$. For the purposes of our investigation, we select $14$ massive clusters with halo masses in the range $10^{13.8-15}~\msun$ and spectroscopic redshifts of $0.867 \lt z  \lt 1.368$. Eleven of these clusters were selected from the Spitzer Adaptation of the Red-sequence Cluster Survey \citep[SpARCS,][]{Wilson09, Muzzin09, Demarco10}, where they were detected in shallow $z'$ and {\it Spitzer}/IRAC $3.6\mu$m images due to their overdensity of red-sequence galaxies \citep{GladdersYee00}. The remaining three clusters were drawn from the South Pole Telescope (SPT) survey \citep{Brodwin10, Foley11, Stalder13} and were initially detected via their Sunyaev-Zeldovich \citep{SunyaevZeldovich70} signature and later spectroscopically confirmed. Table~\ref{table:1} lists the properties of our cluster sample including halo mass ($M_{200}$) and radial scale ($R_{200}$) -- which are both obtained using the MAMPOSSt method \citep{MBB13} as outlined in \citet{Biviano21}.

\begin{table}
\centering
\resizebox{\columnwidth}{!}{\begin{tabular}{cccccc}
    \hline
    \hline  
    
    \multirow{2}{*}{Name}  &  $M_{200}$  & $R_{200}$ & $\sigma$ & \multirow{2}{*}{$z$} & {$N_{\rm members}$} \\ 
    & [$10^{14}~{\rm M}_{\odot}$] & $[\rm cMpc]$ & [km s$^{-1}$]  & & [$>10^{10}~\msun$]\\
    
    \hline

    SpARCS0034 &   0.6 &   1.08 &   $700 \pm 150$   &  0.867 &  23\\
    SpARCS0035 &    3.8 &   2.17 &   $840 \pm 50$    &  1.335 &  18\\
    SpARCS0036 &    3.6 &  2.09 &   $750 \pm 90$   &  0.869 &  45\\
    SpARCS0215 &    2.4 &   1.70 &   $640 \pm 130$    &  1.004 &  34\\
    SpARCS0335 &    1.8 &   1.59 &   $540 \pm 30$    &  1.368 &  7\\
    SpARCS1047 &    2.5 &   1.78 &   $660 \pm 120$    &  0.956 &  26\\    
    SpARCS1051 &    2.2 &   1.80  &   $690 \pm 40$   &  1.035 &  26\\
    SpARCS1613 &    11.1 &   2.97 &   $1350 \pm 100$    &  0.871 &  68\\
    SpARCS1616 &    3.3 &   1.98 &   $780 \pm 40$    &  1.156 &  39\\
    SpARCS1634 &    2.7 &   1.85 &   $715 \pm 40$    &  1.177 &  34\\
    SpARCS1638 &    1.7 &  1.56  &   $565 \pm 30$   &  1.196 &  20\\
    SPT0205    &    3.1 &   1.77 &   $680 \pm 60$   &  1.323 &  19\\
    SPT0546    &    5.8 &   2.42 &   $980 \pm 70$    &  1.067&  27\\
    SPT2106    &    7.3 &   2.62 &   $1055 \pm 85$    &  1.131 &  30\\

    \hline
    \hline
    \label{table:obs_galaxies}
    \label{table:1}
\end{tabular}}
\caption{Properties of our observed cluster sample, including $\mtwo$, $R_{200}$, velocity dispersion, cluster redshift, and the number of spectroscopic members (with $\mstar >10^{10}~\msun$). The values in the $R_{200}$ and $\mtwo$ columns were obtained using the MAMPOSSt method \citep{MBB13} as outlined in \citet{Biviano21}. Details regarding the cluster membership criteria are discussed in \S\ref{subsec:2.2}, while information regarding the total number of members used to measure the velocity dispersion is provided in Table 1 of \citet{Balogh21}.}
\end{table}

\subsection{Cluster Membership and Classification}
\label{subsec:2.2}

We define our initial satellite population to consists of all objects -- excluding the central -- within $R_{200}$ (projected) of a given cluster and with a stellar mass $\mstar > 10^{10}~\msun$ -- i.e. the $\sim 80\%$ stellar mass completeness limit for the photometric sample \citep{vdB20}. In addition, for objects with a secure spectroscopic redshift (Redshift\_Quality\footnote{Please refer to \citet{Balogh21} for a description of the redshift quality flags and the assignment process.} = 3,4), we limit our satellite population to those systems with $\lvert z_{\rm{spec}} - z_{\rm{cluster}} \rvert  \leq 0.02(1+z_{\rm{spec}})$. Meanwhile, for sources without a secure spectroscopic redshift, we define the members of the satellite population as those systems with STAR $\neq$ 1 and $\lvert z_{\rm{phot}} - z_{\rm{cluster}} \rvert  \leq 0.08(1+z_{\rm{phot}}$),  where the STAR flag is the GOGREEN star/galaxy classification based on color selection as described in \cite{vdB20}.
As discussed in \citetalias{Baxter22}, the photometric redshift selection was informed by our knowledge that the $z_{\rm phot}$ uncertainty for galaxies more massive than $10^{10}~\msun$ is $0.048(1+z)$. Nevertheless, we find that if we subsequently characterize and account for interlopers and incompleteness, as described in \S\ref{subsec:2.3}, the results of our analysis do not depend on the $\Delta z$ threshold adopted as part of 
this particular membership criterion. Altogether, these membership selection criteria yield a total of 1072 cluster members (416 spectroscopic/656 photometric). Lastly, we classify the quiescent members of our cluster population using the following rest-frame $UVJ$ color-color cuts defined by \citet[][see also \citealt{Williams09}]{Whitaker11}:

\begin{equation}
\begin{split}
({U-V}) > 1.3~\cap~({V-J}) \lt 1.6~\cap~  \\ 
({U-V}) > 0.88 \times ({V-J}) +  0.59  \; .
\end{split}
\label{uvj_eqn}
\end{equation}

\subsection{Completeness Correction}
\label{subsec:2.3}

Following the methodology utilized in \citet{vdB13, vdB20}, we apply a completeness correction to account for incompleteness and interlopers that contaminate our photometric sample. 
To accomplish this, we compute a membership correction factor based on the subset of galaxies that have both multi-band photometry and spectroscopic redshift measurements, and subsequently apply this factor to the photometric sample.
The membership correction factor (Eqn.~\ref{c_factor}) is defined as the sum of the number of galaxies that are either secure cluster members and false negatives divided by the sum of the number of secure cluster members and false positives, 
\begin{equation}
\label{c_factor}
\textit{C}_{\rm{factor}} = \frac{\textit{N}(\rm{secure~cluster}) + \textit{N}(\rm{false~negative})  }{\textit{N}(\rm{secure~cluster})+ \textit{N}(\rm{false~positive})} \; .
\end{equation}
Secure cluster members are objects with spectroscopic \emph{and} photometric redshifts that are consistent with cluster membership. 
False negatives, on the other hand, refer to objects that are spectroscopically confirmed as cluster members but have photometric redshifts inconsistent with cluster membership.
Conversely, false positives are objects that are not cluster members based on their spectroscopic redshift, yet exhibit photometric redshifts consistent with the redshift of the cluster.

To account for the presumed color dependence of field contamination, we separately compute the correction factor for star-forming and quiescent galaxies. 
Moreover, we compute the correction factor within bins of stellar mass (ranging from $10^{10.0-11.4}~\msun$) and $R_{\rm{proj}}/\rtwo$ (ranging from 0 to 1) for both galaxy populations. 
Notably, we observe a negligible variation in the completeness correction with respect to galaxy color, as the correction factor applied to star-forming and quiescent populations differs by less than $2\%$.

Lastly, we apply the appropriate correction factor as a weight to each cluster member. 
This adjustment leads to a modest change in the measured quenched fractions ($\sim1-2.5\%$). 
Importantly, this completeness correction has no bearing on the final results of our analysis or the conclusions drawn, as they remain consistent irrespective of its application.

\section{Simulated Cluster Sample}
\label{sec:TNG}

\subsection{IllustrisTNG Cluster Sample}
\label{subsec:3.1}

As in \citetalias{Baxter22}, we once again construct our simulated cluster population -- which is matched on redshift to our observed cluster sample -- using the TNG300-1 simulation from the IllustrisTNG project\footnote{\href{https://www.tng-project.org}{https://www.tng-project.org}} \citep[TNG,][]{Nelson18, Niaman18, Springel18, Pillepich18, Marinacci18}. TNG300-1 is a large volume ($\sim300~{\rm cMpc}^{3}$), high-resolution ($2 \times 2500^{2}$ resolution elements), cosmological, gravo-magnetohydrodynamical simulation that utilizes the moving mesh \texttt{AREPO} code and solves for the coupled evolution of dark matter, cosmic gas, luminous stars, and supermassive black holes from a starting redshift of $z=127$ to the present day, $z=0$. TNG300-1 has a dark matter (gas) mass resolution of $m_{\rm DM} = 5.9 \times 10^{7}~\msun$ ($m_{\rm baryon} = 1.1 \times 10^{7}~\msun$), which corresponds to a halo mass (stellar mass) completeness of $\sim10^{10}~\msun$ ($\sim10^{9}~\msun$). As explained in \S3.3 of \cite{Pillepich18}, we augment the stellar masses for TNG300-1 galaxies at $z\sim 1$ by a factor of $1.3\times$ to account for resolution limitations that systematically underestimate stellar masses within the simulations.  

Our simulated cluster population consists of $56$ unique clusters ($\mtwo \gt 10^{14}~\msun$) drawn from 10 snapshots that range from $z=1.36$ to $z=0.85$ with a median redshift of $z=1.1$, where the median redshift difference between an observed cluster and its simulated analog is $|\Delta z| \sim 0.03$. As previously described in \S3.1 of \citetalias{Baxter22}, our simulated cluster population is constructed to match the redshift distribution of our observed cluster sample. We accomplish this by separating both datasets into equal redshift bins and selecting four unique simulated clusters for each observed cluster within a particular bin.

\subsection{Satellite Membership in Simulated Cluster Population}
\label{subsec:3.2}

We apply the exact cluster membership criteria as described in section \S3.2 of \citetalias{Baxter22}; please refer to this work for a more detailed description of our membership selection procedure. 
In short, our simulated satellite population consists of objects that satisfy the following conditions: (\emph{i}) located within $\rtwo$ of a given cluster as measured at the redshift of observation ($z_{\rm obs}$) and (\emph{ii}) objects with resolution-corrected stellar mass of $\mstar > 10^{10}~\msun$ measured at $z_{\rm obs}$ -- where the stellar masses are given by the total mass of all star particles associated with each galaxy (i.e. IllustrisTNG Subhalo-MassType masses with Type=4). Our final simulated cluster population includes 1220 cluster members across the 56 simulated clusters. Though our simulated cluster sample is comprised of more hosts than the observed cluster sample, the former is biased towards less-massive systems ($\mhalo \lt 10^{14.3}~\msun$) – see figure 1 of \citetalias{Baxter22}. However, as explained in \S3.1 of \citetalias{Baxter22}, this bias towards low-mass hosts has a negligible impact on our results due to there being a weak dependence between the distribution of satellite infall times (at fixed stellar mass) and host halo mass. We confirmed this by comparing the cumulative infall time distribution for satellites, at fixed stellar mass, as a function of host halo mass. We observed that, at fixed stellar mass, the infall time distribution for satellites in low-mass and high-mass clusters exhibits only a weak dependence on host mass, with average infall time differences of $\sim 0.02-0.03$ Gyr.

\section{Modeling Environmental Quenching}
\label{sec:Model}

\subsection{Updated Environmental Quenching Model}
\label{subsec:4.1}

In our previous work \citetalias{Baxter22}, we developed an infall-based environmental quenching model to constrain the quenching timescale required to reproduce the satellite quiescent fraction versus satellite stellar mass trend as measured in our aforementioned observed cluster sample. The updated environmental quenching model developed in this investigation shares many similarities with the original model in that it (\emph{i}) accounts for the contribution from \textquote{field quenching} in the simulated cluster population using the coeval field quenched fraction measurements derived from the Cosmic Assembly Near-Infrared Deep Extragalactic Legacy Survey \citep[CANDELS,][]{Grogin11, Koekemoer11, Guo13, Galametz13, Santini15, Stefanon17, Nayyeri17, Barro19} -- for more details, see $\S4.2$ of \citetalias{Baxter22}; (\emph{ii}) incorporates the contributions from satellite pre-processing \citep{Fujita04, DeLucia12, Werner22, Salerno22} in the infall region ($1-3 ~\rtwo$) of the clusters -- for more details, see $\S6.3$ of \citetalias{Baxter22}; and (\emph{iii}) implements an infall-based environmental quenching model in which quenching of the simulated satellites occurs some time $\tau_{\mathrm{quench}}$ after the first crossing of $R_{200}$
-- for more details, see $\S4.3$ of \citetalias{Baxter22}. 
The model proved to be highly successful, reproducing the observed satellite stellar mass function and satellite quenched fraction trends -- i.e. the satellite quenched fraction as a function of stellar mass, projected host-centric radius, and redshift -- associated with our observed cluster population at $z \gtrsim 1$. 
The inferred satellite quenching timescale was found to be mass-dependent and consistent with the empirically-derived cold gas ( H$_2$ + H{\scriptsize I}) depletion timescale at intermediate $z$ from \citet{Popping15}, suggesting that starvation is the dominant quenching mechanism at $z \lt 2$. 

The objective of the investigation herein is to test the validity of the aforementioned conclusion by developing a generalized model for environmental quenching that allows $R_{\mathrm{quench}}$ -- i.e. the radius at which quenching, and therefore the clock measuring $\tau_{\mathrm{quench}}$, is assumed to begin -- to vary as a free parameter.
While some environmental quenching studies use estimates of the virial radius of the host halo (e.g. $\rtwo$) as the physical location at which environmental quenching begins \citep[e.g.][]{Balogh00, Fham18}, it has been found that both cold gas stripping and the removal of diffuse gas from the circumgalactic medium of a galaxy can begin to occur beyond $\rtwo$ \citep{Bahe13, Cen14, Zhang19, Ayromlou21}. 
As mentioned above, our original model accounts for this scenario by allowing quenching to occur in the infall regions ($1$--$3~\rtwo$) of our clusters – see \S6.3 of \citetalias{Baxter22} for a description of how this is implemented in our model. 
Furthermore, certain environmental quenching mechanisms are simply more efficient at smaller host-centric radii – e.g.~RPS is most efficient near pericenter \citep{Cortese21, Boselli22}. 
Therefore, by imposing the condition that $R_{\mathrm{quench}}= 1.0~\rtwo$, the environmental quenching model developed in \citetalias{Baxter22} neglects potentially important regions of parameter space, thereby potentially overlooking alternative quenching pathways. 
The impact of including $R_{\mathrm{quench}}$ as a free model parameter is that, under the assumption that satellite orbits are \emph{not} exclusively radial, it allows the model to potentially explore quenching pathways distinct from the \textquote{special case} assumed in \citetalias{Baxter22}. Should the aforementioned assumption be invalid, our model would suffer from a severe degeneracy between $R_{\rm quench}$ and $\tau_{\rm quench}$, limiting the amount of new information that could be gained from adding the quenching radius as a free parameter. Additionally, distinguishing between a slow quenching process and a long delay time followed by rapid quenching would be challenging. However, if the satellite galaxies exhibit a mix of orbital anisotropies, which perhaps depend on mass and redshift, this degeneracy can be partially broken.


\begin{figure}
 \centering
 \hspace*{-0.25in}
 \includegraphics[width=3.5in]{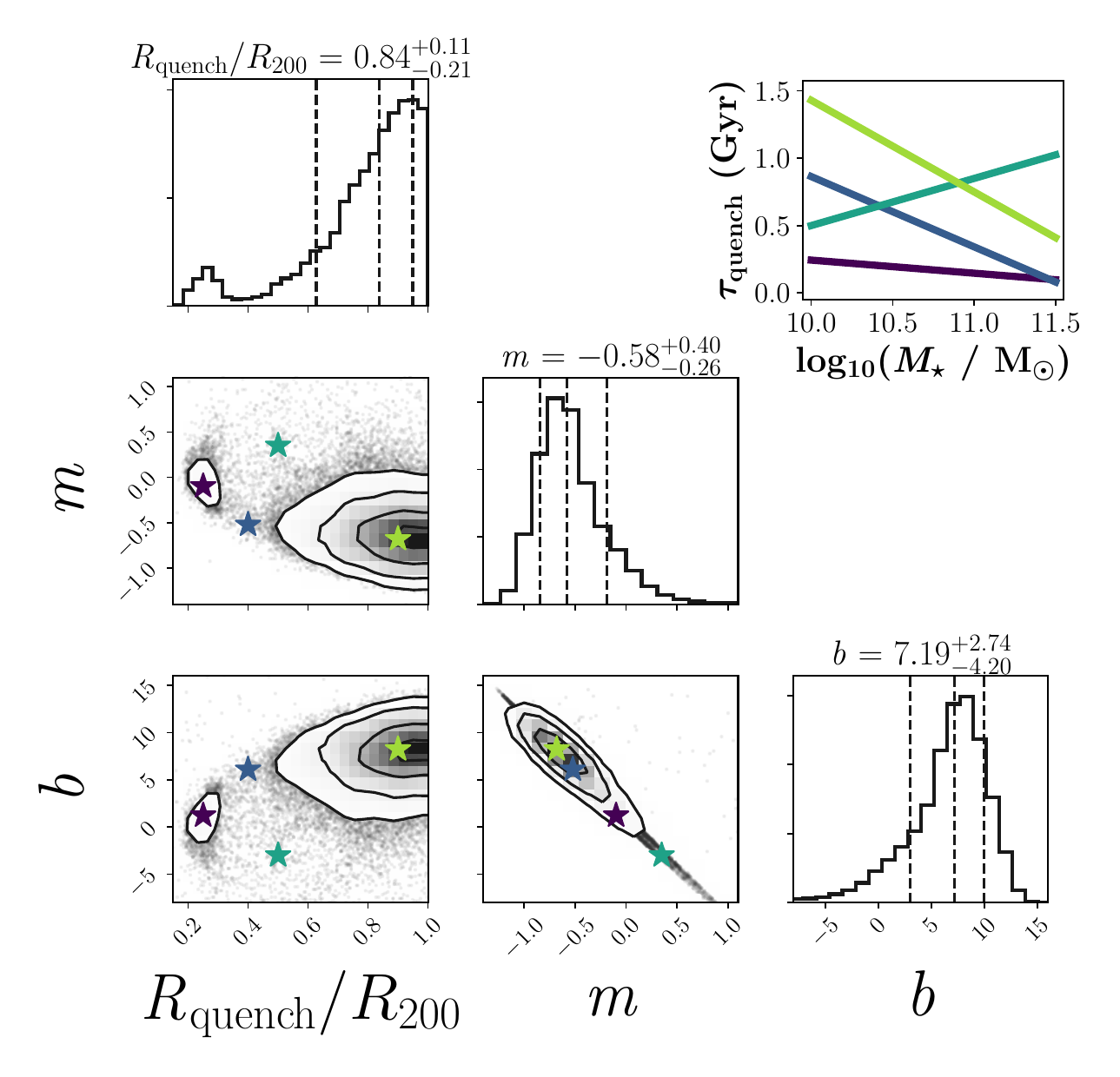}
\caption{Corner plot showing the one- and two-dimensional projections of the posterior probability distributions of the environmental quenching model parameters. The $16^{\rm th}$, $50^{\rm th}$, and $84^{\rm th}$ percentiles associated with the model parameters are shown by the dashed vertical lines. The contours are drawn from the $0.5\sigma$ to $2\sigma$ level in increments of $0.5\sigma$. The model parameters associated with the highest likelihood model are $R_{\mathrm{quench}} = 0.90~\rtwo$, $m=-0.68$, and $b=8.23$, consistent with those found in \citetalias{Baxter22}. The 1D posterior probability distribution of $R_{\mathrm{quench}}$ has an additional local maxima located at $R_{\mathrm{quench}} \sim 0.25~\rtwo$, suggesting that there is another region -- albeit relatively small -- in this parameter space with solutions that are potentially consistent with the observed satellite quenched fraction trends at $z\gtrsim 1$. We test this by isolating four solutions at $0.25, 0.40, 0.50,$ and $0.90~\rtwo$ -- depicted by the filled stars -- and performing a more in-depth analysis of how each reproduces the observations. Lastly, the inset in the top right-hand corner illustrates the quenching timescale associated with each of the aforementioned solutions.}
 \label{fig:fig1}
\end{figure}

\begin{figure*}
\centering
\hspace*{-0.1in}
\includegraphics[width=7.0in]{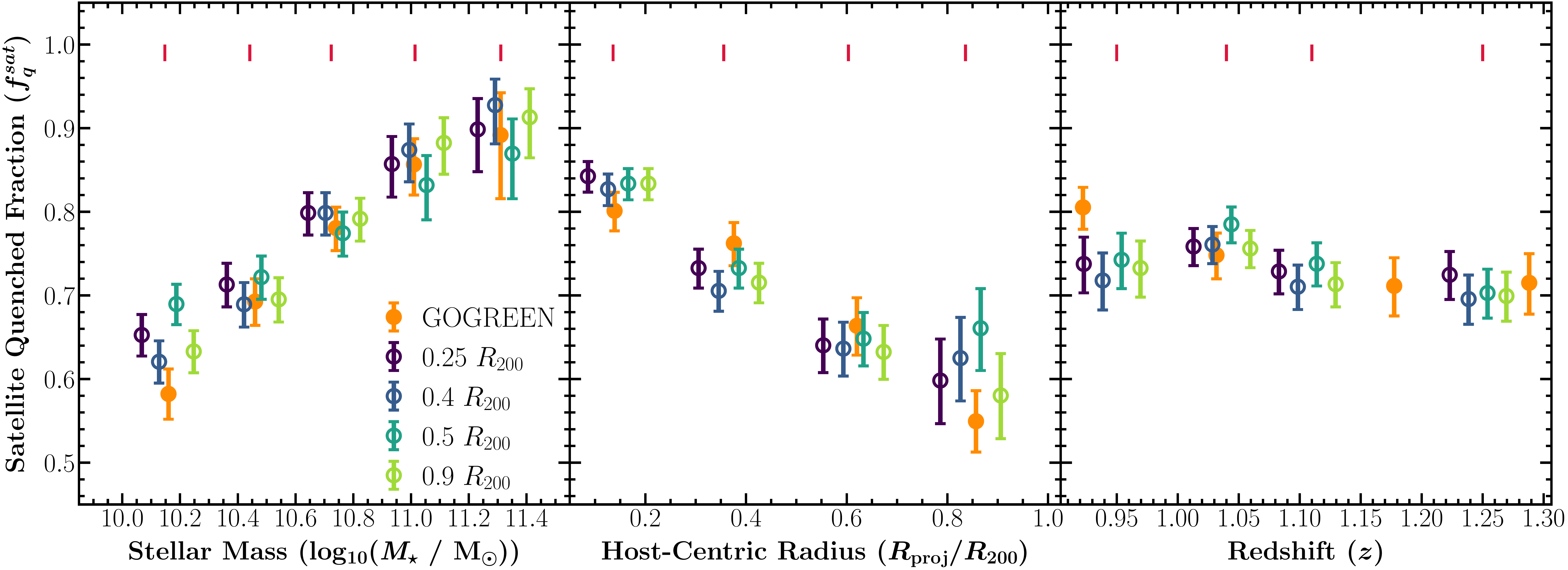}
\caption{Satellite quenched fraction as a function of satellite stellar mass (left panel), projected host-centric radius (middle panel), and redshift (right panel). The orange circles are the measurements associated with our observed cluster sample. The remaining circles are the measurements associated with the four solutions highlighted in Fig.~\ref{fig:fig1}, which are labeled according to their associated $R_{\rm quench}$ values. To enhance clarity, we introduce a slight horizontal offset to visually differentiate between the different models, while also including vertical red lines at the top of each panel to indicate the position of the unaltered values. We note that the median redshift bins between the simulated and observed data are inherently offset, largely due to the former being based on discrete snapshots instead of continuous values. With the exception of the model with $R_{\rm quench}=0.50~\rtwo$, which overproduces the satellite quenched fraction at low masses and large host-centric radius, these results suggest that there exists a broad range of solutions in the $R_{\mathrm{quench}}$-$\tau_{\mathrm{quench}}(\mstar)$ parameter space that yield models able to reproduce the observed satellite quiescent fraction as a function of stellar mass, host-centric radius, and redshift as probed by the GOGREEN data set.}
\label{fig:fig2}
\end{figure*}

%
Another modification is the inclusion of the condition that environmental quenching can only occur at $z \lt 2.5$, so as to allow for the 
formation of a hot halo or dense ICM whereby mechanisms such as starvation and RPS can thereby effectively act to quench cluster members \citep[e.g.][]{Harshan23}. 
In other words, it is difficult to explain how potential environmental quenching mechanisms could effectively operate prior to the emergence of massive, virialized halos with a hot or dense ICM.
In practice, this constraint potentially allows for a small fraction of satellites ($\lesssim 7\%$) that are accreted prior to $z=2.5$ to quench almost immediately after this condition is satisfied. 

Finally, we now also perform a comprehensive Monte Carlo Markov Chain (MCMC) analysis using the \texttt{emcee} ensemble sampler package \citep{Foreman13}. This step is included to ensure that the parameter space associated with the updated environmental quenching model is thoroughly explored, with the primary parameters being the radius at which environmental quenching begins ($R_{\mathrm{quench}}$) along with the slope ($m$) and $y$-intercept ($b$) of the satellite quenching timescale ($\tau_{\mathrm{quench}}$) that we allow to vary linearly with satellite stellar mass, as defined below:
\begin{equation}
\label{eq:tau_quench}
\tau_{\rm{quench}}= m*\log_{10}(\mstar/\msun) + b.
\end{equation}
Given that our model inherently accounts for quenching in the infall region ($1-3~\rtwo$), we limit $R_{\mathrm{quench}} \lt 1.0~\rtwo$. Furthermore, we apply uniform priors to all model parameters and define the log likelihood function as  
\begin{equation}\label{eq:loglike}
    \begin{multlined}
    \ln\,p(y\,|\,R_{\mathrm{quench}},m,b,f ) = \\ - \frac{1}{2} \sum_{i=1}^{N}  \left[ \frac{(y_{i,\rm{obs}} - y_{i, \rm{model}})^{2}}{s_{i}^{2}} + \ln \left ( 2\pi\,s_{i}^2\right) \right] \, ,
    \end{multlined}
\end{equation}
where 
\begin{equation*}
      s_{i}^{2} = \sigma_{i}^{2} + f^{2}y_{i, \rm{model}}^{2} \, .
\end{equation*}
Thus, our chosen likelihood function is a Gaussian where the observed variance ($\sigma_{i}^{2}$) is assumed to be underestimated by a fractional amount $f$ in order to account for the possibility that the uncertainties \textit{are not} Gaussian\footnote{In fact, the uncertainties that correspond to our observed quiescent fractions are binomial, however, we find that in general $\sigma_{\rm lower} \approx \sigma_{\rm upper}$. For this reason, we simply define $\sigma = \sigma_{lower}$.} and uncorrelated. Lastly, $y_{\rm{obs}}$ and $y_{\rm{model}}$ are 1D vectors that contain, respectively, the observed and predicted satellite quenched fractions binned as a function of satellite stellar mass, host-centric radius, and redshift. In the following section, we discuss the results from our Bayesian inference analysis. 

\section{Results}
\label{sec:Results}

\subsection{MCMC Analysis \boldmath$\&$ Competing Solutions}
\label{subsec:5.1}

As a reminder, the two primary parameters of our environmental quenching model are the host-centric radius where quenching begins ($R_{\rm quench}$) and the time -- as measured from $R_{\rm quench}$ -- required for satellites to environmentally quench ($\tau_{\rm quench}$). 
The utility of this model is that by using the infall histories of our simulated satellite population we are able to predict the quiescent fraction as a function of satellite stellar mass, host-centric radius, and redshift. 
Thus, the goal of this Bayesian inference analysis is to determine the model parameters that are most consistent with observed data by comparing model results with the quiescent fraction measurements derived from our observed cluster sample.
Although several initial configurations were tested -- all yielding similar conclusions -- the MCMC results that we discuss herein were acquired using $100$ walkers initialized in a tiny Gaussian ball centered on $R_{\mathrm{quench}}=1.0~\rtwo$, $m=-0.6$, and $b=0.80$.
For this particular configuration, it took $45{,}800$ steps for the model to converge, where the condition for convergence is defined such that the number of steps taken is greater than 100 times the average auto-correlation time.
We find that the highest likelihood model occurs when $R_{\mathrm{quench}} = 0.90 ~\rtwo$, $m=-0.68$, and $b=8.23$, whereas the $16^{\rm th}$, $50^{\rm th}$, and $84^{\rm th}$ percentiles of the model parameters are given by $R_{\mathrm{quench}} = 0.84\substack{+0.11 \\ -0.21}~\rtwo$, $m=-0.58\substack{+0.40 \\ -0.26}$, and $b=7.19\substack{+2.74 \\ -4.20}$. 

The results from our Bayesian inference analysis are summarized in Fig.~\ref{fig:fig1}, displaying a corner plot that depicts the 1D and joint 2D posterior probability distributions of our model parameters.
The first notable observation is that there exist two local maxima in the marginalized distribution of $R_{\rm quench}$ (top-left panel of Fig.~\ref{fig:fig1}) at $\sim 0.25$ and $1.0~\rtwo$, respectively. 
This suggests that there are non-unique solutions in the parameter space of our environmental quenching model that are potentially consistent with observations.
However, the relative importance of these two local maxima implies that the potential solutions associated with the less prominent peak are confined to a more limited region within the model parameter space.
The second notable observation is that there is a well-defined \textquote{ridge} of quenching timescales, as illustrated by the strong covariance between the slope and $y$-intercept of the linear satellite quenching timescale (bottom row, middle column of Fig.~\ref{fig:fig1}). 
Specifically, this ridge shows that there are three classes of quenching timescales that are potentially permissible according to our environmental quenching model (see the inset in the top-right corner of Fig.~\ref{fig:fig1}). 
The first class consists of quenching timescales that decrease with increasing satellite stellar mass -- i.e. the region with $m \lt 0$. 
The second class consists of short quenching timescales that are largely independent of satellite stellar mass -- i.e. the region around $m \sim 0$.
The third class consists of quenching timescales that increase with increasing satellite stellar mass -- i.e. the region with ${m} \gt 0$. 
Interestingly, despite the highest likelihood model being found at $R_{\mathrm{quench}} = 0.90~\rtwo$, the aforementioned observations suggest that the $R_{\rm quench}$-$\tau_{\rm quench}$ parameter space is potentially degenerate with a range of possible solutions that are consistent with observations. 
This observation aligns with the recent findings from \citet{Tacchella22}, which indicate that galaxies likely undergo quenching over a diverse range of timescales.
Moreover, recent studies have also highlighted a similar degeneracy between the onset of quenching and the quenching timescale at $z\sim0$ \citep{Oman21, Reeves23}, signaling the need for additional observable(s) beyond the quiescent fraction to constrain these parameters.

To investigate whether this degeneracy is present in our environmental quenching model, we select four solutions in our model parameter space -- illustrated by the four colored stars in Fig.~\ref{fig:fig1} -- and directly compare their estimated quiescent fractions with observations. 
These solutions are selected to probe specific regions of our model's parameter space -- i.e. the two local maxima (purple, light-green), the 
\textquote{saddle} between the local maxima (blue), and the outskirts of the covariance relationship between the slope and y-intercept of the linear quenching timescale (green). 
Furthermore, these four solutions, henceforth denoted by their respective $R_{\rm quench}$ values, are purposely selected to run the gamut of potentially permissible classes of quenching timescales (see top-right inset in Fig.~\ref{fig:fig1}). 
As shown in Fig.~\ref{fig:fig2}, nearly all of these solutions are \textit{roughly} consistent with the observed satellite quenched fraction trends as a function of satellite stellar mass, host-centric radius, and redshift. 
The only exception occurs for the solution that probes the outskirts of the covariance between the slope and the $y$-intercept of the linear quenching timescale, given that it overpredicts the quiescent fraction at both low satellite stellar mass and large host-centric radius by more than $2\sigma$. 
This suggest that quenching timescales that increase towards higher satellite stellar mass are inconsistent with observations, and as such, we will no longer consider the $m \gt 0$ family of solutions. 

In summary, we find that our results are degenerate given that there exists a region in the multi-dimensional parameter space in which seemingly distinct solutions return quiescent fraction trends that are consistent with observations. Nevertheless, two additional questions naturally arise from this observation with the first one being: (\emph{i}) is it possible to rule out solutions by comparing their results with additional measurements derived from our observed cluster population?; and (\emph{ii}) are these seemingly disparate solutions truly distinct \textit{or} do they represent the same underlying quenching pathway? Regarding the first question, we address it immediately in the following subsection, however, we save discussion of the second question for \S\ref{subsec:6.2}.

\subsection{Comparison of the Observed and Estimated Properties of Transition Galaxies}
\label{subsec:5.2}

One feasible approach to break the aforementioned degeneracy would be to use information from the environmental quenching model -- e.g. the time at which satellites environmentally quench -- to isolate a population of \textquote{transition galaxies} and compare their properties with observations.
By \textquote{transition galaxies}, we are specifically referring to a population of galaxies that are currently in the process of quenching their star formation \emph{or} have recently completed this process. 
From an observational standpoint, we define the former to include massive galaxies ($\gt 10^{10}~\msun$) in the \textquote{green valley} \citep[GV,][]{Schiminovich07, Schawinski14, Vulcani15}, while the latter is defined to include massive galaxies classified as \textquote{post-starbursts} \citep[PSB,][]{D&G83, D&G92, C&S87}.
We isolate these galaxies in our observed cluster sample following the selection criteria described in Table 2 of \citet{McNab21}, where massive GV galaxies are defined by their position in rest-frame ($NUV-V$) and ($V-J$) colour-colour space \citep{Moutard16, Moutard16b, Moutard18, Leja19} and massive PSBs are selected based on their D4000 and $\oii$ spectral indices \citep{Muzzin14}.
Thus, this definition includes galaxies with an assortment of quenching timescales and pathways, which we explore further in \S\ref{subsec:6.1}.

\begin{figure}
 \centering
 \hspace*{-0.25in}
 \includegraphics[width=3.5in]{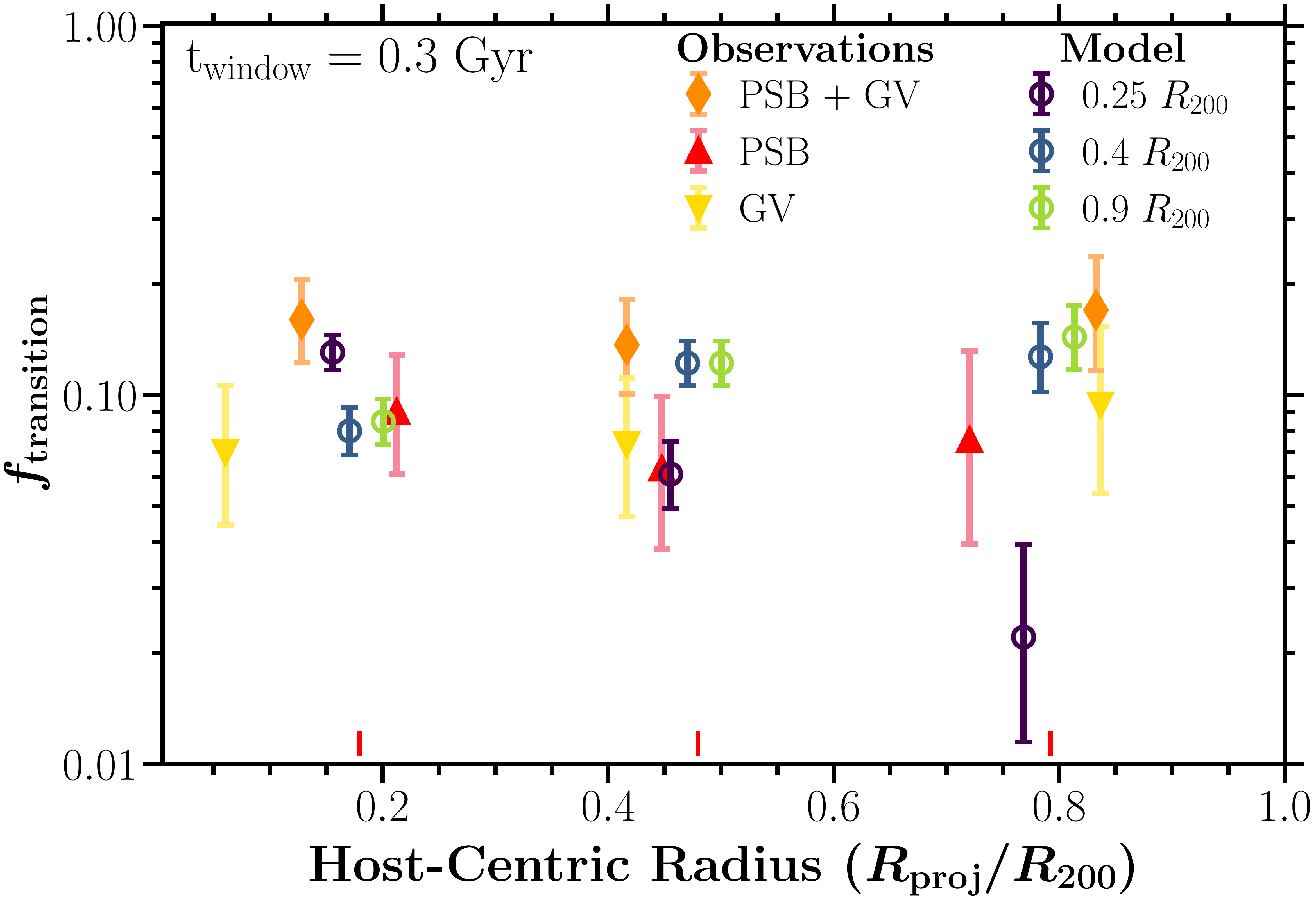}
 \caption{The relative abundance of transition galaxies as a function of host-centric radius. The orange diamond show the combined relative abundance of massive ($\gt 10^{10}~\msun$) PSB and GV galaxies from the GOGREEN cluster sample \citep{McNab21}, whereas the red and yellow triangles depict these abundances separately. The open circles depict the model results in which the simulated transition galaxy population is defined to include both massive galaxies that have either recently quenched (i.e., PSB analogs) or are in the process of quenching (i.e., GV analogs). In particular, for a given model we define the transition population as satellites that either quench $\lt 0.30$ Gyr before $t_{\mathrm{obs}}$ \textit{or} star-forming satellites that will quench $\lt 0.30$ Gyr after $t_{\mathrm{obs}}$. To improve clarity, a small horizontal offset is applied to distinguish between the various models, while red vertical lines are included slightly above the horizontal axis to mark the position of the unaltered values. We find that all models -- with the exception of those at $R_{\rm quench} = 0.25~\rtwo$ -- are generally consistent with the combined observed abundance of GV and PSB galaxies.}
 \label{fig:fig3}
\end{figure}
The open circles represent the model outcomes based on the assumption that the transition population includes galaxies that have either recently quenched or are currently undergoing quenching

\begin{figure}
 \centering
 \hspace*{-0.25in}
 \includegraphics[width=0.5\textwidth]{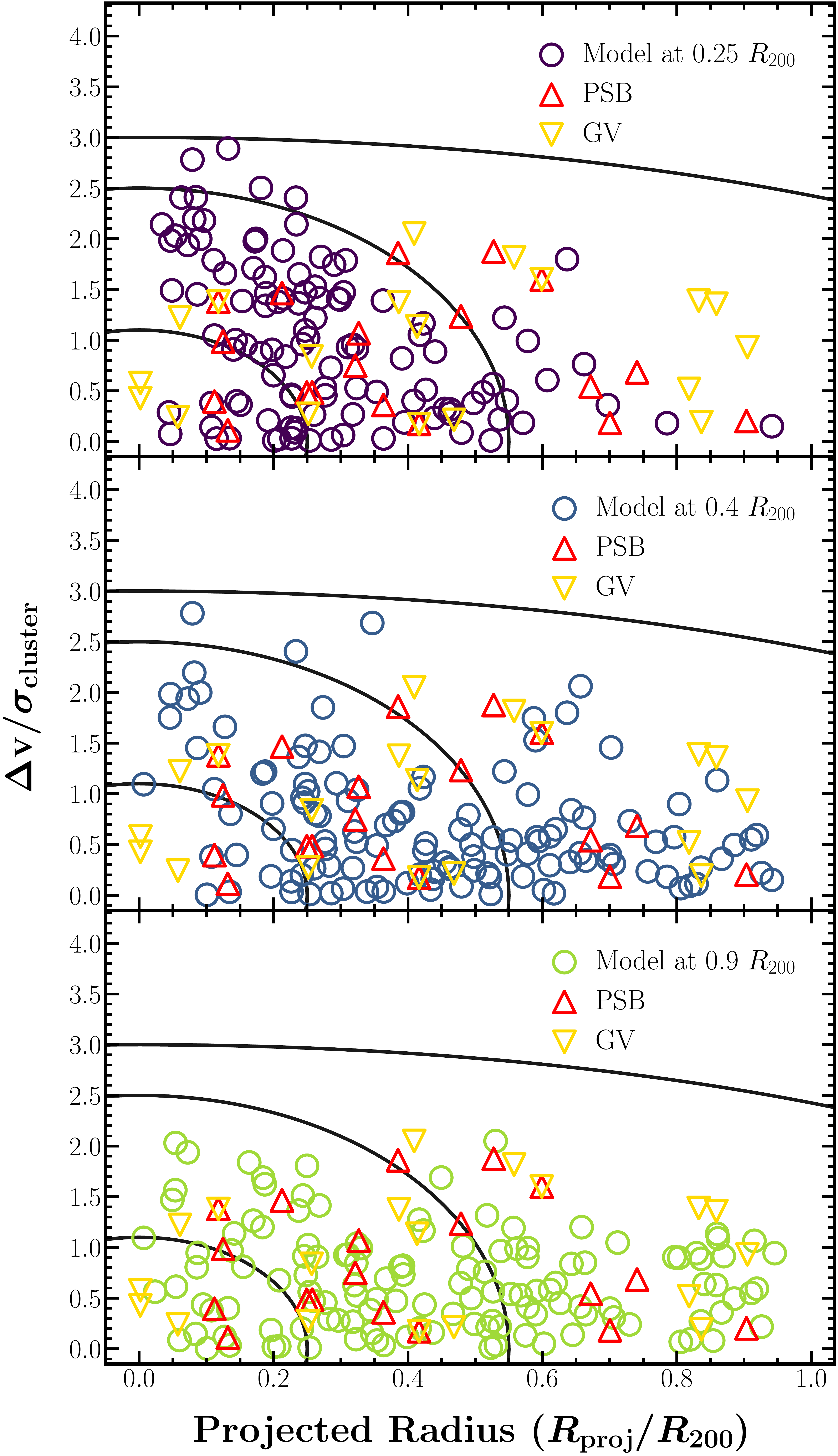}
 \caption{The ``folded'' projected phase-space distribution for transition galaxies selected at the redshift of observation. Each panel compares the projected phase-space distribution associated with the transition galaxies selected from one of the three solution drawn from our environmental quenching model at $R_{\rm quench} = 0.25$, $0.4$, and $0.9~\rtwo$ with the corresponding distribution of massive PSB (red triangles) and GV (yellow triangles) galaxies identified in the GOGREEN cluster sample. The solid contours illustrate the phase-space bins adopted by \citet{Muzzin14}. We observe that the solution at $R_{\rm quench}=0.25~\rtwo$ has a relative dearth of transition galaxies in the outer regions of the cluster. Moreover, within the inner $\lesssim 0.30-0.35~\rtwo$, the $R_{\mathrm{quench}} = 0.25~\rtwo$ solution yields transition galaxies with much higher line-of-sight velocities relative to the observed transition galaxy population. A similar argument could also be made for the $R_{\mathrm{quench}} = 0.40~\rtwo$ solution, such that only the $R_{\mathrm{quench}} = 0.90~\rtwo$ solution yields line-of-sight velocities in the inner regions of the cluster that are roughly consistent with observations.}
 \label{fig:fig4}
\end{figure}


\begin{figure}
 \centering
 \hspace*{-0.25in}
 \includegraphics[width=3.5in]{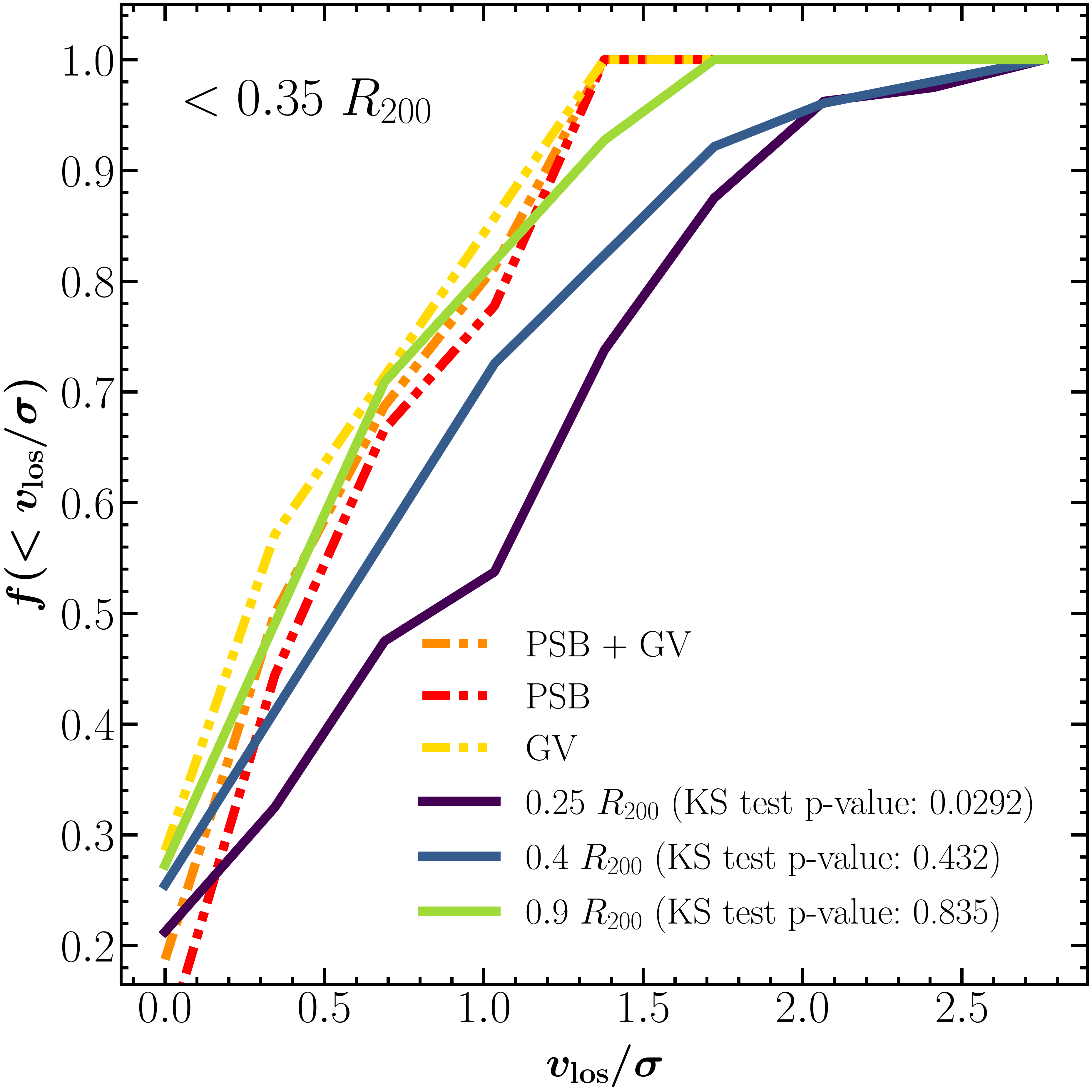}
 \caption{The cumulative distribution of the line-of-sight velocities for transition galaxies located within the inner projected $0.35~\rtwo$ of their host. The dotted-dashed orange line depicts the combined distribution of massive PSB and GV galaxies, whereas the red and yellow dotted-dashed lines show their separate distribution. All of the other lines correspond to the distributions derived from the competing environmental quenching models. The legend includes the KS two-sided $p$-values derived from comparing the combined observed and modeled line-of-sight velocity distributions. This analysis shows that the null hypothesis can only be rejected for the model with $R_{\rm quench} = 0.25~\rtwo$.}
 \label{fig:fig5}
\end{figure}


Our approach to isolate the population of transition galaxies associated with our environmental quenching models is to assume that these galaxies are only visible for a limited time window, t$_{\rm window}$, relative to the redshift of observation of our simulated cluster sample.
This definition is inspired by the concept of the \textquote{visibility time} of transition galaxies, which refers to the limited period during which the defining features of transition galaxies, such as intermediate colors and strong Balmer absorption lines, can be observed. 
In the framework of our model, we identify the transition galaxy population as galaxies that are within $\pm 0.30$ Gyr of quenching, as measured relative to the redshift of observation. Specifically, transition galaxies are those that satisfy either of the following conditions:
\begin{enumerate}[labelindent=0pt,labelwidth=\widthof{\ref{last-item}},label=\roman*.,itemindent=1em,leftmargin=!]
\item quiescent at $t_{\rm obs}$ $\wedge~t_{\rm q} < t_{\rm obs} + 0.30$  Gyr.
\item star forming at $t_{\rm obs}$ $\wedge~t_{\rm q} > t_{\rm obs} - 0.30$ Gyr.\label{last-item}

\end{enumerate}
Here, $t_{\rm q}$ is the lookback time where quenching concludes defined as $t_{\rm q} = t_{\rm cross} - \tau_{\rm quench}$, where $t_{\rm cross}$ is the lookback time at which a galaxy crosses $R_{\rm quench}$ and $\tau_{\rm quench}$ is the satellite quenching timescale. 
As will be discussed in \S\ref{subsec:6.1}, this is consistent with the timescales associated with various classes of observationally-identified transition galaxies --- e.g. massive ($\gt 10^{10}~\msun$) PSB and GV galaxies. 
In Fig.~\ref{fig:fig3}, we compare the relative abundance of transition galaxies for each quenching model relative to the abundance of massive GV and PSB galaxies identified in the GOGREEN cluster sample from \citet{McNab21}.
We find that the quenching model with $R_{\rm quench}=0.25~\rtwo$ underproduces the observed relative abundance of transition galaxies beyond the very inner regions of the cluster (mainly due to relatively rapid quenching timescale and small quenching radius).
Meanwhile, the other two models (with $R_{\rm quench}=0.4~\rtwo$ and $R_{\rm quench}=0.9~\rtwo$) are generally consistent with the observed abundance of transition galaxies as a function of host-centric distance. 

In Fig.~\ref{fig:fig4}, we also compare the projected phase-space distribution of the simulated transition galaxies with the observed distribution of transition galaxies, as constrained by massive PSB and GV galaxies in the GOGREEN sample. 
The first notable observation, in line with the results from Fig.~\ref{fig:fig3}, is that the solution at  $R_{\rm quench}=0.25~\rtwo$ yields very few transition galaxies in the outer regions of the cluster.
Additionally, within the inner $\lesssim 0.30-0.35~\rtwo$, the $R_{\mathrm{quench}} = 0.25~\rtwo$ solution yields transition galaxies with much higher line-of-sight velocities relative to the observed transition galaxy population. 
On the surface, it appears that only the $R_{\mathrm{quench}} = 0.90~\rtwo$ solution yields line-of-sight velocities in the inner regions of the cluster that are roughly consistent with observations. 
To test this, we compute the cumulative line-of-sight velocity normalized by the cluster velocity dispersion ($v_{\mathrm{los}}/\sigma$) distributions of the inferred transition galaxies, limited to the inner $0.35~\rtwo$ of the cluster, and compare the results with the corresponding distribution for the observed sample of transition galaxies from GOGREEN. 
This information is shown in Fig.~\ref{fig:fig5} along with the Kolmogorov-Smirnov (KS) two-sided $p$-values. 
The first major takeaway is that, in addition to failing to reproduce the observed relative abundance of transition galaxies, the model at $R_{\rm quench}=0.25~\rtwo$ yields a $p$-value less than $0.05$, indicating that the null hypothesis can be rejected – i.e. the transition population predicted by this model is not drawn from the same parent distribution as the observed sample. 
Consequently, we consider the solution at $R_{\rm quench}=0.25~\rtwo$ to be less viable as it does not adequately reproduce the observed relative abundance of transition galaxies and results in an overabundance of high-velocity satellites in the inner regions of the cluster.
Lastly, these results imply that only the solutions with relatively long and mass dependent timescales are unable to be rejected based on the KS test.
This, in turn, brings us back to the second question posed at the end of \S\ref{subsec:5.1} – i.e. do these solutions represent the same quenching pathway with apparent differences driven by a covariance between $\tau_{\rm quench}$ and $R_{\rm quench}$?  
In addition to addressing this question, in the following section \S\ref{sec:Discussion}, we explore how the aforementioned conclusion depends on our definition of transition galaxies as well as how our results compare with previous environmental quenching studies at $z \sim 1$.

\section{Discussion}
\label{sec:Discussion}

\subsection{Transition Galaxies and Visibility Times}
\label{subsec:6.1}

As mentioned in \S\ref{subsec:5.2}, our approach for isolating the population of transition galaxies within the framework of our environmental quenching model assumes that these galaxies are visible for a limited time window, t$_{\rm window}$, relative to the redshift of observation in our simulated cluster sample. 
For PSB galaxies, the visibility times typically indicates the time required for the galaxy's Balmer absorption lines to weaken to the level of a quiescent galaxy, often inferred from the equivalent width measurement of the H${\delta}$ absorption line in the galaxy's spectrum. 
In contrast, for green valley (GV) galaxies, the visibility time (referred to as the \textquote{crossing time}) signifies the time required to cross the green valley and is typically inferred using statistical analyses of galaxy properties in the GV region of the color-magnitude diagram.

Studies have indicated that PSB galaxies have a relatively short visibility time, with estimates ranging from $0.1-1$ Gyr \citep{Wild09, Muzzin14, Wild16, French18, Rowlands18, Belli19, Wild20}. 
On the other hand, GV galaxies have a more extended visibility time, with some studies suggesting that the transition phase can last up to $1-2$ Gyr \citep{Bremer18, Forrest18, Smethurst18, Noirot22}. 
Moreover, as shown in \citet{Moutard16b} the visibility time of GV galaxies depends on stellar mass such that low-mass ($\lt 10^{9.5}~\msun$) galaxies tend to follow a fast quenching channel ($\sim 0.4$ Gyr to cross the GV) to become PSB while high-mass ($\gt 10^{10}~\msun$), evolved galaxies follow a slow quenching channel ($1-3.5$ Gyr to cross the GV). 
Furthermore, an investigation by \citet{Schawinski14} found that this timescale depends on morphology, with early-type galaxies crossing the green valley in timescales of less than $0.25$ Gyr, and late-type galaxies crossing it in less than $1.0$ Gyr. 
Given that our study is limited to massive ($\gt 10^{10}~\msun$) GV galaxies, we caution readers against extrapolating our results to lower stellar masses.
Overall, the exact duration of the visibility time for PSB and GV galaxies depends on various factors, including the methodology for identifying them, the specific diagnostic used to estimate transition timescales, the spectral resolution, the signal-to-noise ratio of the observations, and other observables such as the host environment \citep{Paccagnella17, Paccagnella19, Socolovsky19, Mao22} and galaxy mass \citep{McNab21}. 
Therefore, the exact visibility of PSB and GV galaxies is influenced by multiple factors, making it challenging to determine precisely. However, for the purposes of this analysis, our chosen visibility time is selected to encompass both recently quenched galaxies and those that are on the verge of quenching, as defined relative to the redshift of observation.

\begin{figure}
\centering
\hspace*{-0.25in}
\includegraphics[width=3.5in]{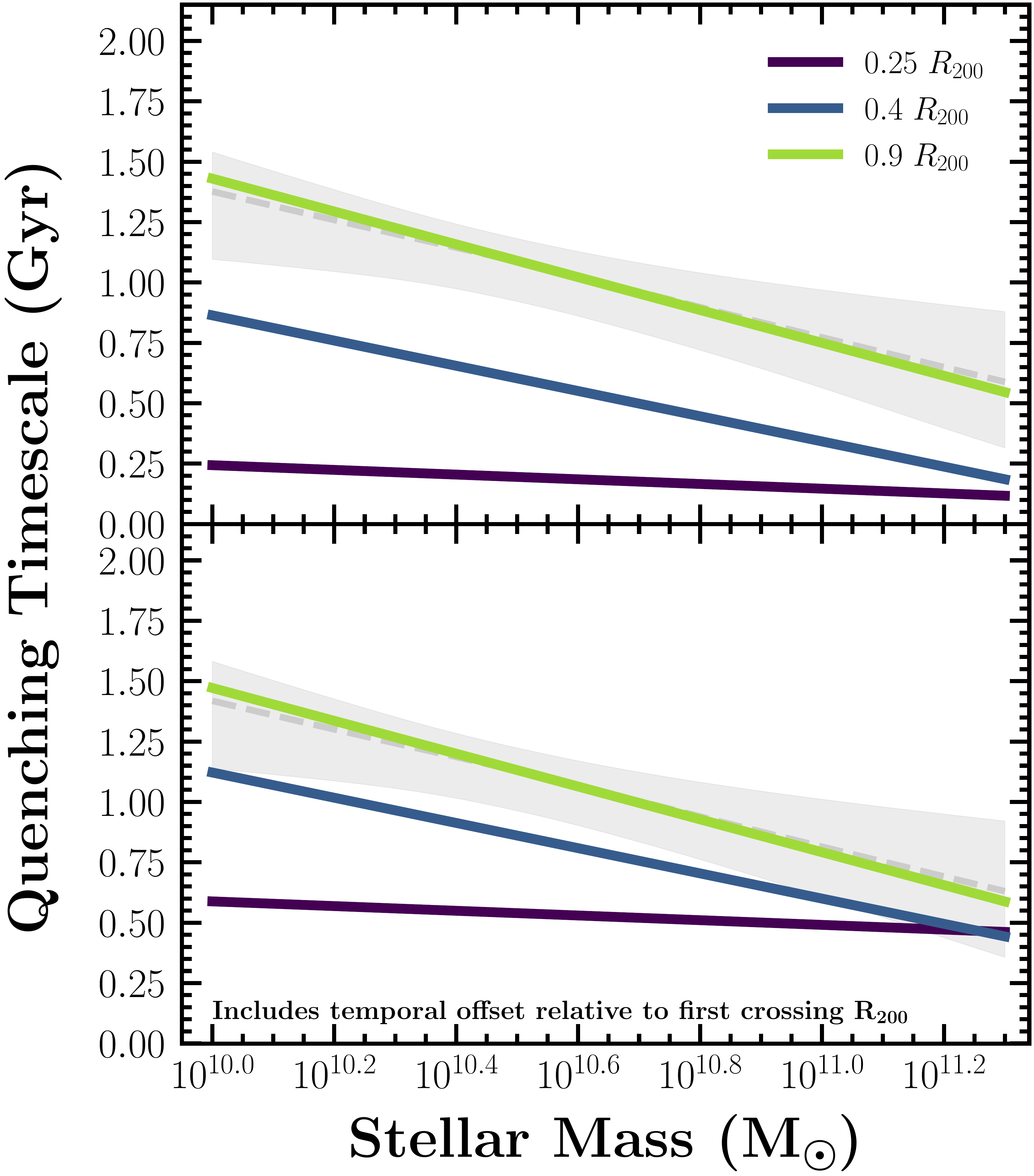}
\caption{Quenching timescales versus satellite stellar mass. The colored lines correspond to the three observationally consistent solutions to our environmental quenching model isolated in Fig.~\ref{fig:fig1}. The dashed grey line and shaded band represent the results associated with the median and corresponding 1-sigma error of the model parameters derived from our MCMC analysis. The upper panel shows the quenching as measured from the time of crossing $R_{\rm quench}$, whereas the lower panel augments these timescales by adding the median time required for the satellites in a given model to travel from $1.0~\rtwo$ to $R_{\rm quench}$. These results, namely that the timescales associated with the various solutions do not overlap after taking into consideration the delay time between first crossing $1.0~\rtwo$ and reaching $R_{\rm quench}$, suggest that our satellite orbits are not exclusively radial.}
\label{fig:fig6}
\end{figure}

\begin{figure*}
 \centering
 \hspace*{-0.1in}
 \includegraphics[width=7.0in]{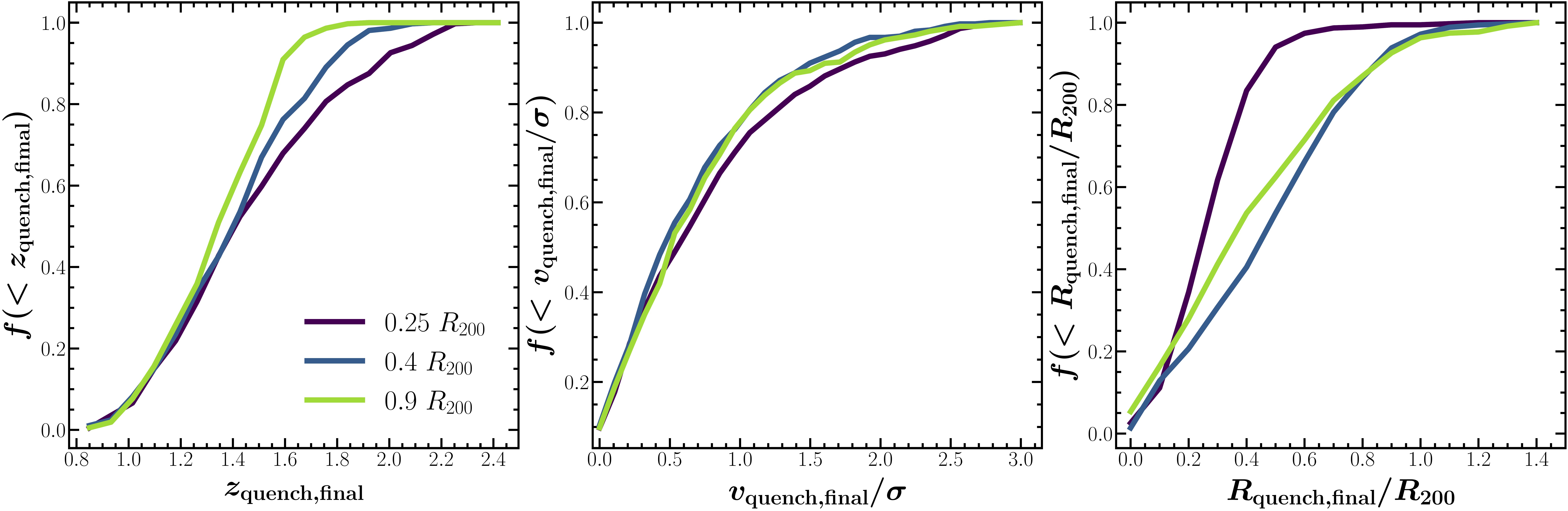}
 \caption{The cumulative distribution for the redshift (left panel), velocity (middle panel), and host-centric radius (left panel) corresponding to the time at which the competing models fully quenched their satellite population. With the exception of the models with $R_{\rm quench} \leq 0.25~\rtwo$, all of the models have strongly overlapping cumulative distributions for the radius \textit{and} line-of-sight velocity at which their satellite populations were environmentally quenched. However, for redshifts above $z \sim 1.3$, there is a clear stratification of the models such that the fraction of galaxies that quench at earlier times increases as $R_{\rm quench}$ decreases. These results suggest that the models with $R_{\rm quench} \leq 0.25~\rtwo$ experience a distinct quenching pathway from the other models given that they environmentally quench the bulk of their galaxies at earlier times, smaller host-centric radii, and with relatively higher line-of-sight velocities.}
 \label{fig:fig7}
\end{figure*}

The selection of transition galaxies in this analysis, namely satellites within $\pm 0.30$ Gyr of quenching as measured relative to $t_{\rm obs}$, aligns with the aforementioned estimates of the visibility time for massive PSB and GV galaxies – i.e. $0.1-1$ Gyr and $0.25-2$ Gyr, respectively. 
While this choice is consistent with observations, widening the visibility window, for example, to $\pm 0.60$ Gyr, would result in the $R_{\rm quench} = 0.25~\rtwo$ solution predicting a relative abundance of transition galaxies consistent with observations. 
However, the second conclusion regarding the $R_{\rm quench} = 0.25~\rtwo$ solution, namely an overabundance of transition galaxies with high line-of-sight velocities in the inner region of the cluster, remains true and even worsens if the visibility window is expanded. 
Likewise, we find that the general conclusion drawn in \S\ref{subsec:5.2} remains true, namely that only solutions with long quenching timescales ($\tau_{\mathrm{quench}} \gtrsim 1$ Gyr) and large quenching radius ($R_{\rm quench} \gtrsim 0.4~\rtwo$), are capable of reproducing the observed abundance \emph{and} phase-space distribution of transition galaxies in clusters at $z \sim 1$. 
This holds even if we modify our definition of observed transition galaxies to only include recently quenched galaxies (i.e., massive PSB) \emph{or} galaxies in the process of quenching  (i.e., massive GV).
Considering the similar relative abundances and projected phase-space distributions of both galaxy populations (as shown in Figs. \ref{fig:fig3} and \ref{fig:fig4}), we choose to combine them to enhance statistical robustness at the cost of defining a population with an assortment of visibility times.

\subsection{Distinct Quenching Pathways?}
\label{subsec:6.2}

As stated in \S\ref{subsec:5.1}, it is important to determine if the various observationally consistent solutions truly represent distinct environmental quenching mechanisms \textit{or} if instead they represent the same quenching mechanism with the differences in quenching timescales being directly tied to changes in the host-centric radius at which quenching begins.
A simple method to test this is to compare the quenching timescale results associated with each of the solutions, which we show in Fig.~\ref{fig:fig6}. 
The top panel depicts the quenching timescales relative to crossing $R_{\rm quench}$, whereas the bottom panel augments this timescale by adding the median time required for a satellite to travel from $1.0~\rtwo$ to $R_{\rm quench}$. 
The results from Fig.~\ref{fig:fig6} suggest that despite having different assumptions for where quenching begins, the solutions at $0.40$ and $0.90~\rtwo$ yield fairly consistent quenching timescales when measured relative to $1.0~\rtwo$. 
Moreover, as illustrated in Fig.~9 from \citetalias{Baxter22}, the timescales associated with these two solutions are roughly consistent with the empirically-derived cold gas ( H$_2$ + H{\scriptsize I}) depletion timescale at intermediate $z$ from \citet{Popping15}. 
Following the logic presented in that analysis, we interpret these solutions to potentially be associated with starvation as the dominant quenching pathway. 
Nevertheless, additional information is required to determine if the solution at $R_{\rm quench}=0.25~\rtwo$ represents a distinct quenching pathway.

A more detailed method of testing if these solutions represent distinct quenching pathways is to compare the properties of their satellite populations – e.g. positions and velocities – at the time in which the quenching process ends. 
Thus, we compare the cumulative distributions of the host-centric radius at the time in which the three solutions fully environmentally quench their satellite population ($R_{\mathrm{quench, final}}$) along with the corresponding line-of-sight velocity and redshift ($v_{\mathrm{quench, final}}/\sigma$ and $z_{\mathrm{quench, final}}$, respectively). 
Together with the quenching timescale information, these additional constraints allow us to answer the following questions: (\emph{i}) how long does the satellite quenching process last? (\emph{ii}) where in the cluster does satellite quenching begin and end?; (\emph{iii}) what is the velocity distribution of satellites at the moment at which quenching ends?

The left, middle, and right panels in Fig.~\ref{fig:fig7}, respectively, compare the cumulative distributions of $z_{\mathrm{quench, final}}$, $R_{\mathrm{quench, final}}$, and $v_{\mathrm{quench, final}}/\sigma$ associated with each of the solutions. 
The first notable observation is that $z_{\mathrm{quench, final}}$ depends on $R_{\mathrm{quench}}$ such that the solutions for which quenching begins at larger (smaller) radii finish quenching at later (earlier) times. 
In line with the results shown in the bottom panel of Fig.~\ref{fig:fig6}, this indicates that for models with a small $R_{\rm quench}$, the time interval between becoming a satellite (i.e., first crossing $\rtwo$) and reaching $R_{\mathrm{quench}}$ is \textit{shorter} than the time required to quench satellites for the models with a large $R_{\rm quench}$.
Additionally, we observe that the solutions at $0.40$ and $0.90~\rtwo$ have consistent cumulative distribution of $R_{\mathrm{quench, final}}$ and $v_{\mathrm{quench, final}}/\sigma$. 
This suggests that these solutions are agnostic towards where quenching begins given that they quench their satellite populations at similar host-centric radii and with overlapping line-of-sight velocities distributions. 
By the same token, we observe that the quiescent satellites associated with the solution at $R_{\mathrm{quench}} = 0.25~\rtwo$ predominantly quench in the core of the cluster ($80\%$ quenched at $\lt 0.40~\rtwo$) with relatively high line-of-sight velocities.

We interpret the results from Figs.~\ref{fig:fig6} and \ref{fig:fig7} as evidence of two distinct quenching pathways, which we define as \textquote{starvation} and \textquote{core-quenching}. 
The former, which applies to the solutions with $R_{\mathrm{quench}} = 0.40$ and $0.90~\rtwo$, is characterized by relatively long ($\gt 1.0$ Gyr) mass-dependent quenching timescales that are roughly consistent with the total cold gas (H$_{2}$+H{\scriptsize I}) depletion timescale at intermediate $z$. 
Meanwhile, the latter is characterized by satellites with relatively high line-of-sight velocities that quench on short timescales ($\sim 0.25$ Gyr) after reaching the inner region of the cluster ($\lt 0.25~\rtwo$).
It is interesting to note that the \textquote{core-quenching} pathway and RPS exhibit similar characteristics: both tend to quench high-velocity satellites located at small distances from their host galaxy's center, and the quenching occurs relatively quickly ($\lesssim 1$ Gyr) \citep{Boselli22}. 
These similarities raise the possibility that the \textquote{core-quenching} pathway could be similar to the RPS mechanism responsible for forming \textquote{jellyfish galaxies} \citep{Poggianti17, Vulcani20a}, especially since many of these galaxies are also observed in the inner regions of clusters ($\lt 0.40~\rtwo$) \citep{Gullieuszik20}.
Nevertheless, while the idea is captivating, we assert that it is beyond the scope of this study to establish a direct equivalence between the \textquote{core-quenching} pathway and RPS.

\begin{figure*}
 \centering
 \hspace*{-0.1in}
 \includegraphics[width=7.0in]{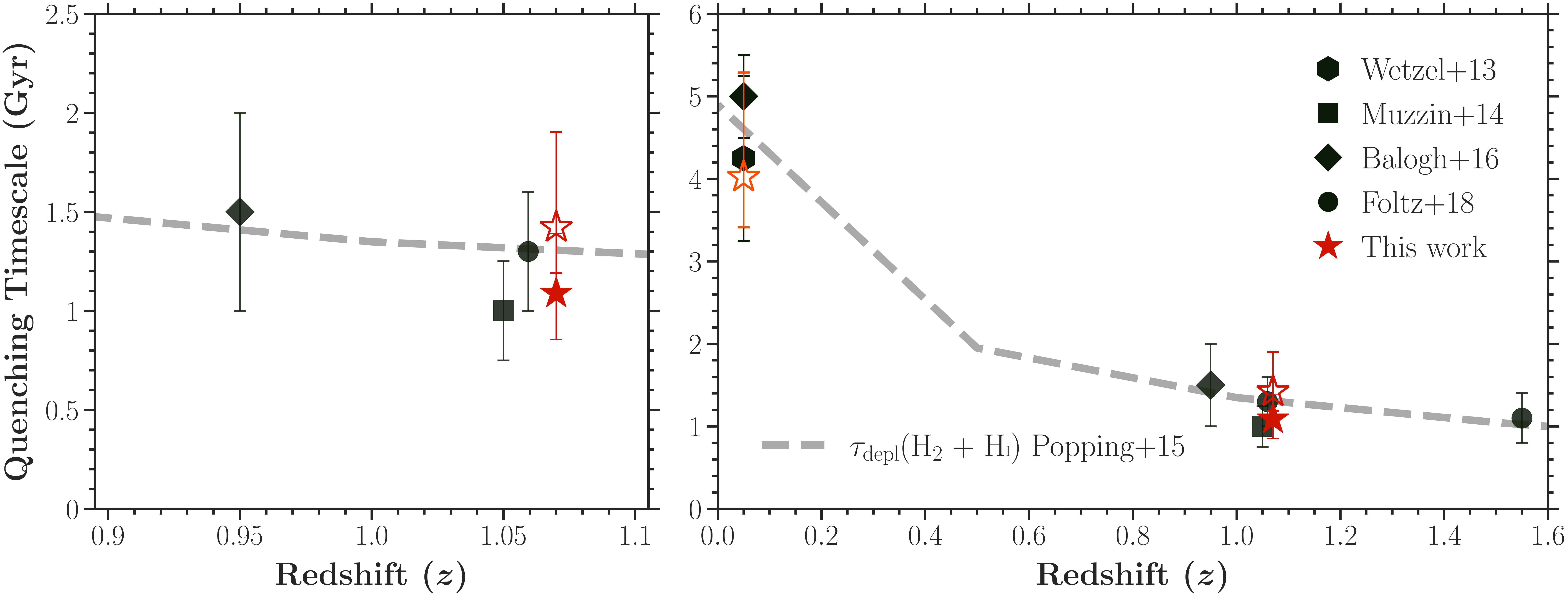}
 \caption{Quenching timescale versus redshift for satellites of massive clusters ($\mhalo \sim 10^{14-15}~\msun$). The filled (unfilled) red star represents the quenching timescale measured at $\mstar = 10^{10.5}~\msun$ ($\mstar = 10^{10}~\msun$) derived from our MCMC analysis (i.e. the dashed grey line in Fig.~\ref{fig:fig6}). Likewise, the orange unfilled star represents the quenching timescale measured at $\mstar = 10^{10}~\msun$ scaled according to the evolution of the dynamical time – $\tau_{\mathrm{quench}}(\mstar) \times (1 + z)^{-1.5}$. The black points show the quenching timescales obtained from comparable studies of environmental quenching in clusters at $z \sim 1$ (left panel) and $0 \lt z \lt 1.6$ (right panel) as measured by \citet{Wetzel13}, \citet{Muzzin14}, \citet{Balogh16}, and \citet{Foltz18}. With the exception of the point from \citet{Wetzel13}, which is evaluated at $\mstar = 10^{10.0}~\msun$,  all of the results from other studies are evaluated for satellites with $\mstar \gt 10^{10.5}~\msun$. The dashed gray line depicts the empirically-derived cold gas ( H$_2$ + H{\scriptsize I}) depletion timescale from \citet{Popping15} evaluated at $\mstar = 10^{10.5}~\msun$.}
 \label{fig:fig8}
\end{figure*}

\subsection{Comparison with Previous Studies}
\label{subsec:6.3}

In the left panel of Fig.~\ref{fig:fig8}, we compare the quenching timescale inferred from this investigation with results from previous environmental quenching studies of cluster populations ($\mhalo \gt 10^{14}~\msun$) at $z \sim 1$ for satellites with $\mstar \gt 10^{10.5}~\msun$.
These studies include \citet{Muzzin14}, \citet{Balogh16}, and \citet{Foltz18}, and they were selected given that they utilize a compatible definition of $\tau_{\rm quench}$ – i.e. defined as the timescale upon which satellites quench as measured relative to first infall.   
Nevertheless, we acknowledge that these studies utilize distinct methodologies for inferring the quenching timescale. 
For example, \citet{Balogh16} inferred quenching timescales of $1.5 \pm 0.5$ Gyr by relating the passive fraction in 10 galaxy clusters from the GCLASS survey to infall histories estimated from semi-analytic simulations. 
Meanwhile, \citet{Muzzin14} used galaxy spectral features to identify PSB galaxies in the GCLASS cluster sample and related the distribution of this population in phase space to the phase-space distribution of infalling subhalos in dark-matter-only zoom-in simulations to obtain a quenching timescale of $1.0 \pm 0.25$ Gyr.
Lastly, \citet{Foltz18} inferred a total quenching timescale of $1.3 \pm 0.3$ by relating the observed numbers of star-forming, quiescent, and green valley galaxies in 10 galaxy clusters to a simulated cluster mass accretion rate using a \textquote{delayed-then-rapid} quenching model \citep{Wetzel13, McGee14, Mok14, Balogh16, Fossati17}.

Despite the different methodologies utilized in these studies, the inferred timescales broadly agree that satellite quenching at $z \gtrsim 1$ proceeds on timescales between $1-1.5$ Gyr following accretion onto an established cluster. 
As shown in Fig.~\ref{fig:fig8}, these timescales are all roughly consistent with the total cold gas depletion timescale at this epoch, suggesting that the consumption of cold gas in absence of cosmological accretion -- i.e. starvation -- could be the dominant quenching mechanism at this epoch. 
Nevertheless, it is important to acknowledge the findings of \citet{Muzzin14}, whose PSB-focused quenching study concludes that RPS is the dominant mechanism in massive clusters. 
Likewise, the results of \citet{Foltz18} suggest that quenching takes place on the dynamical timescale of the cluster, although they cannot dismiss the possibility of quenching due to gas depletion in the absence of cosmological accretion.

In the left panel of Fig.~\ref{fig:fig8}, we explore the redshift dependence of the satellite quenching timescale by including results from \citet{Wetzel13} at $z\sim0$ – evaluated at $\mstar=10^{10}~\msun$ for $\mhalo=10^{14-15}~\msun$ – and results at $z \sim 1.6$ from \citet{Foltz18}.
We also include the quenching timescale estimate at $z\sim0$ from \citet{Balogh16}, obtained by scaling $\tau_{\mathrm{quench}}$ according to the dynamical time – i.e. $\tau_{\mathrm{quench}} \times (1 + z)^{-1.5}$.
We perform a similar scaling using our inferred quenching timescale evaluated at $\mstar=10^{10}~\msun$ to obtain an estimate of the quenching timescale at $z\sim0$.
As noted in several previous studies, we find that the satellite quenching timescale evolves roughly like the dynamical time \citep{Tinker10, Balogh16, Foltz18, Baxter22}.
Although the catalyst behind the redshift evolution of the quenching timescale remains unknown, one possible interpretation of the aforementioned observation is that the environmental quenching mechanism(s) responsible for producing the observed quenched fraction results in clusters at $z\sim1$ are potentially equivalent to those at play in their low-$z$ descendants, where the differences in timescales between the separate epochs is due to the evolution of the host system properties (e.g. halo masses, velocity dispersion, etc.), but not the quenching mechanism itself.

In comparing our investigation to previous studies, it is important to highlight that the transition galaxy phase space analysis detailed in \S\ref{subsec:5.2} shares similarities with the approach used in \citet{Muzzin14} to constrain $R_{\rm quench}$ and $\tau_{\rm quench}$ at $z \sim 1$.
Specifically, in \citet{Muzzin14} they compare the projected phase space distribution of PSB galaxies from the GCLASS cluster sample with that of simulated PSB galaxies, which they infer by isolating galaxies in time-steps of $0.2$ Gyr after first passage of $0.25$, $0.50$, and $1.0~\rtwo$. 
Moreover, by using a 2D KS test to compare these distributions, they rule out all scenarios in which quenching begins after the first passage of $1.0~\rtwo$ and lasts between $0.5-1.1$ Gyr. 
Additionally, they find that the combination of $R_{\rm quench}=0.50~\rtwo$ and $\tau_{\rm quench}=1.0$ Gyr is most consistent with the data. 
In essence, despite the fact that the clusters explored in \citet{Muzzin14} constitute a subset of our sample, we arrive at contrasting conclusions regarding where within the cluster - and for how long - quenching takes place.

Nevertheless, comparing these two investigations objectively presents challenges due to several key differences. 
Firstly, these studies employ different populations of transition galaxies and clusters. 
For instance, in the study by \citet{Muzzin14}, the transition galaxies are exclusively limited to spectroscopically-selected PSB galaxies primarily sourced from two massive clusters at a redshift of $z=0.87$. 
In contrast, our study includes both massive photometrically-selected GV galaxies and spectroscopically-selected PSB galaxies as part of the transition galaxy population, predominantly obtained from higher-redshift clusters. 
Considering that previous studies have found evidence for the existence of different quenching channels among distinct observed galaxy populations \citep[e.g.,][]{Whitaker12, Schawinski14, Moutard16, Moutard18}, our observed transition galaxy sample is likely far more heterogeneous in terms of quenching timescales \emph{and} quenching pathways.
Additionally, considering that \citet{Muzzin14} attributes quenching to RPS, one possible interpretation is that the influence of RPS becomes more pronounced with increasing halo mass and decreasing redshift.
Additionally, besides utilizing distinct infall histories to select our simulated transition galaxy populations, both studies employ unique methodologies. 
For example, in contrast to their analysis, we investigate the stellar mass dependence of environmental quenching and track the self-quenching of the infalling field population. 
The consideration of stellar mass dependence is important since, as demonstrated in Figure 4 of \citetalias{Baxter22}, the observed quiescent fraction trends cannot be replicated under the assumption that the quenching timescale, measured since first passage of $1.0~\rtwo$, is independent of satellite stellar mass. 
Moreover, incorporating a self-quenching prescription based on measurements of the observed field quenched fraction introduces an additional stellar mass dependence to the satellite quenching process, indicating that more massive ($\gt 10^{10.5}~\msun$) satellites, on average, would be expected to undergo quenching in the field or infall region compared to their less massive counterparts. 
Considering all these factors, it is highly likely that the discrepancies between our analyses stem from a combination of disparate methodologies and variations in cluster and transition galaxy populations.

In a broader context, this work supports the notion that environmental quenching of massive ($\mstar \gt 10^{10}~\msun$) galaxies operates over a diverse range of timescales and likely involves multiple contributing mechanisms, with starvation playing a significant role. Recent environmental quenching recent studies, such as \citet{Cortese21} and \citet{Alberts22}, have also pointed towards the involvement of multiple mechanisms in quenching satellite galaxies. This also aligns with recent findings presented in \citet{Tacchella22}, which demonstrate that at $z\sim0.8$, massive galaxies in diverse environments exhibit a broad range of quenching timescales and potentially quenching pathways. Conversely, in the context of low-mass galaxies ($\mstar \lt 10^{9.5}~\msun$), research by \citet{Moutard18} has revealed that in densely populated regions of the universe, quiescent galaxies are primarily PSB or recently-quenched galaxies, which suggests a more limited range of quenching timescales in such environments. Altogether, this suggests that quenching processes affecting massive galaxies are complex and multifaceted, with multiple mechanisms at play, while low-mass galaxies appear to undergo quenching through a more uniform process. However, more comprehensive studies that explore quenching over a broad range of redshifts, environments, and stellar masses are required to verify this picture.

\section{Summary and Conclusions}
\label{sec:Conclusion}

In our recent paper, \citetalias{Baxter22}, we investigated the dominant quenching mechanism in massive clusters at $z \gtrsim 1$, using a simple infall-based environmental quenching model parameterized by the quenching timescale $\tau_{\rm quench}$. The success of this model was that it: (\emph{i}) improved upon previous studies by implementing a prescription for field quenching and pre-processing in the infall region; (\emph{ii}) is fairly simple in that it involves one primary parameter - i.e. the satellite quenching timescale $\tau_{\rm quench}$; (\emph{iii}) roughly reproduces the observed satellite stellar mass function as well as the satellite quenched fraction as a function of stellar mass (by construction), host-centric radius, and redshift; and (\emph{iv}) yields quenching timescales that are consistent with the total cold gas depletion time at intermediate $z$, suggesting that \textquote{starvation} - i.e. the depletion of cold gas in the absence of cosmological accretion - is the dominant driver of environmental quenching at $z \lt 2$. 

Thus, the motivation for this follow-up investigation was to further test the validity of this conclusion by developing a more generalized environmental quenching model that allows for potentially distinct quenching pathways through the introduction of the parameter $R_{\rm quench}$ -- i.e. the host-centric radius corresponding to the onset of environmental quenching.
To this end, we performed a comprehensive MCMC analysis to fully explore the parameter space of our updated environmental quenching model, and ultimately discovered two local maxima at approximately $0.25$ and $1.0~\rtwo$ in the 1D posterior probability distribution of $R_{\rm quench}$. 
From here, we isolated four distinct solutions in the $R_{\rm quench}-\tau_{\rm quench}$ parameter space - i.e. two near the aforementioned local maxima, one in the \textquote{saddle} between the local maxima, and one in the outskirts of the covariance relationship between the slope and $y$-intercept of the linear quenching timescale. 
We discovered that, with the exception of the solution in the outskirts of the aforementioned covariance relation, all solutions reproduce the satellite quenched fraction trends associated with our GOGREEN cluster population.

In an effort to determine if these solutions represent distinct quenching pathways, we compared their quenching timescales (relative to first crossing $\rtwo$) as well as their positions and velocities at the time of quenching. Based on this information, we separated the solutions between those driven by \textquote{starvation} and \textquote{core-quenching}. 
The former quenching pathway corresponds to model solutions that exhibit quenching timescales that are aligned reasonably well with the total cold gas (H$_{2}$+H{\scriptsize I}) depletion timescale at intermediate $z$. 
On the other hand, the latter pathway, which bears resemblance to ram-pressure stripping, is characterized by satellites with relatively high line-of-sight velocities, experiencing rapid quenching within a short timescale ($\sim 0.25$ Gyr) after entering the inner region of the cluster ($\lt 0.30~\rtwo$).
To break the degeneracy among these solutions, we compared our model results with observed properties of transition galaxies in massive clusters at $z \gtrsim 1$ from the GOGREEN survey. From this analysis, we found that only the solutions associated with the starvation quenching pathway are consistent with both the observed quiescent fraction trends \textit{and} the phase-space distribution and relative abundance of transition galaxies at $z \gtrsim 1$. 

In conclusion, this investigation provides further insight into the dominant quenching mechanisms in massive clusters at $z \gtrsim 1$, and shows that results from a simple environmental quenching model can be used to isolate distinct quenching pathways. 
By comparing model results with observations, we found that the \textquote{core-quenching} pathway is not consistent with the observed transition galaxy trends.
Conversely, our results are consistent with the scenario in which galaxies quench on relatively long timescale between $1.0-1.5$ Gyr after accretion, thus supporting the idea that starvation may be the dominant quenching mechanism at $z \lt 2$. 
Nonetheless, despite the concordance between the inferred quenching timescales and the total gas depletion time during this epoch, this study provides evidence supporting the importance of group pre-processing in shaping the observed quiescent fraction, as well as the notion that RPS contributes as a secondary mechanism for quenching in massive clusters at $z\gtrsim1$, in line with recent environmental quenching reviews \citep{Cortese21, Alberts22}.

\section*{acknowledgements} 
DCB thanks the LSSTC Data Science Fellowship Program, which is funded by LSSTC, NSF Cybertraining Grant $\#$1829740, the Brinson Foundation, and the Moore Foundation; participation in the program has greatly benefited this work.
MCC and DCB acknowledge support from the National Science Foundation through grant AST-1815475. GHR gratefully acknowledges support from the NSF from grant AST-1517815 and AST-2206473, and from HST programs GO-15294, AR-14310, and NASA ADAP-80NSSC19K0592.
RD gratefully acknowledges support by the ANID BASAL project FB210003. 
FS acknowledges support from a CNES postdoctoral fellowship
GW gratefully acknowledges support from the National Science Foundation through grant AST-2205189 and from HST program number GO-16300. Support for program numberGO-16300 was provided by NASA through grants from the Space Telescope Science Institute, which is operated by the Association of Universities for Research in Astronomy, Incorporated, under NASA contract NAS5-26555.

This research made extensive use of {\texttt{Astropy}},
a community-developed core Python package for Astronomy
\citep{Astropy13, Astropy18}.
Additionally, the Python packages {\texttt{NumPy}} \citep{numpy},
{\texttt{iPython}} \citep{iPython}, {\texttt{SciPy}} \citep{SciPy20}, and
{\texttt{matplotlib}} \citep{matplotlib} were utilized for our data
analysis and presentation. 
In addition, this research has made use of NASA’s Astrophysics Data System Bibliographic Services.
Finally, this work makes use of observations taken by the CANDELS Multi-Cycle Treasury Program with the NASA/ESA HST, which is operated by the Association of Universities for Research in Astronomy, Inc., under NASA contract NAS5-26555.

\section*{Data availability} 
Data sharing is not applicable to this article as no new data were created or analyzed in this study.  

\bibliography{citations}

\section*{Affiliations}

\noindent {\it $\!\!^1$Department of Physics \& Astronomy, University of California, Irvine, 4129 Reines Hall, Irvine, CA 92697, USA \\
$\!\!^{2}$Department of Physics and Astronomy, University of Waterloo, Waterloo, ON N2L 3G1, Canada \\
$\!\!^{3}$Waterloo Centre for Astrophysics, University of Waterloo, Waterloo, ON N2L 3G1, Canada \\
$\!\!^{4}$Department of Physics \& Astronomy, University of Kansas, 1251 Wescoe Hall Drive, Malott room 1082, Lawrence, KS 66045 \\ 
$\!\!^{5}$INAF - Osservatorio Astronomico di Trieste, via G.B. Tiepolo 11, 34143 Trieste, Italy \\
$\!\!^{6}$Departamento de Astronom\'ia, Facultad de Ciencias F\'isicas y Matem\'aticas, Universidad de Concepci\'on, Concepci\'on, Chile \\
$\!\!^{7}$Department of Physics, University of Helsinki, Gustaf H\"allstr\"omin katu 2a, FI-00014 Helsinki, Finland \\
$\!\!^{8}$Department of Physics and Astronomy, University of California Davis, One Shields Avenue, Davis, CA, 95616, USA \\
$\!\!^{9}$Department of Physics and Astronomy, York University, 4700 Keele St., Toronto, Ontario, M3J 1P3, Canada\\
$\!\!^{10}$IRAP, Institut de Recherche en Astrophysique et Planétologie, Université de Toulouse, UPS-OMP, CNRS, CNES,\\ 14 avenue E. Belin, F-31400 Toulouse, France\\
$\!\!^{11}$INAF - Osservatorio astronomico di Padova, Vicolo Osservatorio 5, I-35122 Padova, Italy \\
$\!\!^{12}$Department of Physics \& Astronomy, University of California, Riverside, 900 University Avenue, Riverside, CA 92521, USA \\
$\!\!^{13}$Steward Observatory and Department of Astronomy, 933 N. Cherry Ave, University of Arizona, Tucson, AZ, 85721 
}

\label{lastpage}
\end{document}